\documentclass[prl,aps,amssymb,twocolumn,showpacs,superscriptaddress]{revtex4}

\usepackage{graphicx}

\begin{document}

\title{Evidence for Two Time Scales in Long SNS Junctions}

\author{F.~Chiodi}

\author{M.~Aprili}

\author{B.~Reulet}

\affiliation{Laboratoire de Physique des Solides, UMR8502, b\^atiment 510, Universit\'e Paris-Sud, 91405 ORSAY Cedex, France}

\date{\today}

\begin{abstract}

We use microwave excitation to elucidate the dynamics of long superconductor / normal metal / superconductor  Josephson junctions. By varying the excitation frequency in the range 10 MHz - 40 GHz, we observe that the critical and  retrapping currents, deduced from the dc voltage vs. dc current characteristics of the junction, are set by two different time scales. The critical current increases when the ac frequency is larger than the inverse diffusion time in the normal metal, whereas the retrapping current is strongly modified when the excitation frequency is above the electron-phonon rate in the normal metal. Therefore the critical and retrapping currents are associated with elastic and inelastic scattering, respectively.

\end{abstract}

 \maketitle

Despite the large amount of work devoted to superconductor / normal metal / superconductor (SNS) Josephson junctions, starting from the pioneering work of J. Clarke \cite{pioneer}, their dynamics remains an open issue. The main raison is that the time dependence of the phase difference between the two superconductors, well understood in tunnel junctions, interferes here with that of the out-of-equilibrium quasi-particles in the normal metal. For instance, the origin of the retrapping current $I_r$, the current below which the junction switches back from the dissipative to the non dissipative state, is still under debate \cite{Angers,Courtois}.

One way to probe the junction dynamics is to measure its dc response to an ac excitation. By varying the frequency of the excitation, the relevant time scales can be revealed. A BCS superconductor driven out-of-equilibrium by microwave radiation experiences increased superconductivity. This counterintuitive phenomenon is known as Dayem-Wyatt effect \cite{Dayem,Wyatt} , first observed in microbridges, and then in thin films and strips \cite{films, tj, tj2, strips, strips2}. It is based on the fact that the BCS gap $\Delta$ is related in a self-consistent way to the energy distribution of the quasiparticles $f(\varepsilon)$ by: $1=\lambda\int_0^{+\infty} \rho(\varepsilon) [1-2f(\varepsilon)] d\varepsilon \label{gapeq}$ where $\lambda$ measures the strength of the electron-phonon interaction and $\rho(\varepsilon)$ is the BCS density of states, which depends on $\Delta$. The thermodynamic gap $\Delta(T)$ corresponds to the equilibrium Fermi-Dirac statistics at temperature $T$; driving the system out-of-equilibrium with an ac current modifies the energy distribution of the excitations, and thus the gap. To put the quasiparticles significantly out of equilibrium clearly demands enough power, but also enough speed: inelastic processes tend to restore equilibrium, and as a consequence enhanced gap is observed only when the excitation frequency $f$ is greater than the inverse of the dominant inelastic process, electron-phonon in usual superconductors.  From the above equation, $\Delta(T)$ is found to be fairly insensitive to $T$ at low temperature whereas it varies strongly close to the critical temperature $T_c$. This reflects the sensitivity of $\Delta(T)$ to a small change in the distribution function, and thus it translates into the fact that microwave pumping is effective only very close to $T_c$, experimentally for $T$ within of few percent of $T_c$ \cite{strips}.

Hybrid structures made of superconductors and normal metals offer a rich extension of the physics of superconductivity. In the SNS geometry, due to coherent Andreev reflections at the N/S interface, a finite gap $\tilde\Delta$, the so-called mini-gap,  develops inside the normal metal. In a way similar to the BCS gap equation, $\tilde\Delta$ is determined by the Andreev bound states as well as by the distribution function of the quasiparticles. It is thus natural to expect an enhanced induced superconductivity by ac irradiation \cite{Mercereau,Warlaumont}. However, two facts make the physics of SNS samples more complex than that of superconductors: i) the distribution functions involve not only the inelastic time but also the diffusion time $\tau_D$ of quasiparticles along the N part, since Andreev pairs diffuse from the N/S interfaces to the center of the N part; ii) the density of states is affected by the rf field through the phase of the Andreev pairs being modified by the ac vector potential.

We report an experimental investigation of the influence of a microwave excitation on both the retrapping current $I_r$ and the critical current $I_c$, the maximum supercurrent that can flow through the junction.

We observe that a microwave of high enough frequency $f>f_c$ enhances $I_c$. We relate $f_c$, temperature independent, to the inverse diffusion time through the normal wire. On the other hand, the switching back to the superconducting state is modified when $f>f_r$, where $f_r$ is of the order of the electron-phonon scattering rate $\tau_{e-ph}^{-1}$ in the normal metal, and has the same, strong temperature dependence $f_r\simeq \tau_{e-ph}^{-1}\propto T^3$ ($f_r\ll f_c$ in our temperature range).

\emph{The samples --}
The samples were fabricated by double angle deposition through a suspended mask, formed by a bilayer of 400 nm PES and 60 nm Si$_3$N$_4$. After drawing the mask by e-beam lithography, we etched with a SF$_6$ RIE the Si$_3$N$_4$ layer and then the PES with a O$_2$ plasma. The undercut is about $0.5\;\mu$m \cite{samplefab}. The junctions were made by depositing a first layer of Al (9 nm), here the normal metal, and a second, overlapping layer of Nb (50 nm), the superconductor. The deposition was made at a base pressure of $10^{-9}$ mbar. The critical temperature of the Nb film ranges from 6.8 to 8 K depending on the evaporation conditions.

We have measured six samples of two different lengths: short samples (simply called 'S' below) with length $L=330$ nm, width $w=300$ nm and normal state resistance $R_n=6\;\Omega$, and long samples ('L'), with $L=780$ nm, $w=320$ nm and $R_n= 14.5\;\Omega$. We deduce the diffusion coefficient $D\simeq$50cm$^2s^{-1}$, and the diffusion times $\tau_D=L^2/D$, with $\tau_D^{-1}\simeq8.2$ GHz for the long samples and 46 GHz for the short samples. The different samples of the same length gave similar results. All the measurement were performed above 1.4 K when the Al is normal. We observe no evidence for superconducting fluctuations in Al above this temperature \cite{craven}.

\begin{figure}

\includegraphics[width=\columnwidth]{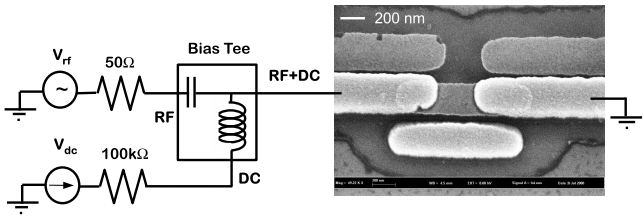}

\caption{Left: Schematics of the experimental setup; Right: SEM picture of sample S.}

\label{fig_setup}

\end{figure}

\emph{Experimental setup --}
The experimental setup is sketched in Fig. \ref{fig_setup}. The samples are dc current-biased through a $100$ k$\Omega$ resistor connected, after filtering, to the dc port of a bias tee, whereas a microwave generator is connected to the rf port. The bias tee is essential for this experiment because it provides a wideband, good coupling of the sample to the rf source. The sample is placed at the end of a $50\;\Omega$ stainless steel coaxial line and sits in the liquid of a pumped He$^4$ cryostat. The temperature is varied  by adjusting the bath pressure. When it is superconducting, the sample has a very small impedance, thus the rf generator of output impedance $R_0=50\Omega$ acts as a current source. This is not strongly modified when the sample is in the normal state since its resistance is $\sim0.1-0.3R_0$. At high frequency the sample has a complex impedance that can be of order $R_0$\cite{Ibias}. However, it is wire bonded to the microstrip on the sample holder. We have measured the inductance in series with the sample, $L=3.3$ nH,  by reflectometry in the range 0.1-5 GHz. This is the dominant impedance at high frequency ($L\omega\sim200\;\Omega$ at 10 GHz), so the sample is still current-biased at high frequency, but the ac current is reduced by a factor $R_0/(L\omega)$ that we take into account, together with the attenuation of the cables that we have measured separately, to estimate the rf power at the sample level. We observe several regimes depending on the frequency of the microwave: at low frequency the voltage follows adiabatically the dc measured $V(I)$ curve; at intermediate frequency, the $V(I)$ is modified and the measured critical current is lowered by the presence of the ac excitation; at high enough frequency we observe a strong enhancement of the critical current.

\begin{figure}

\includegraphics[width=\columnwidth]{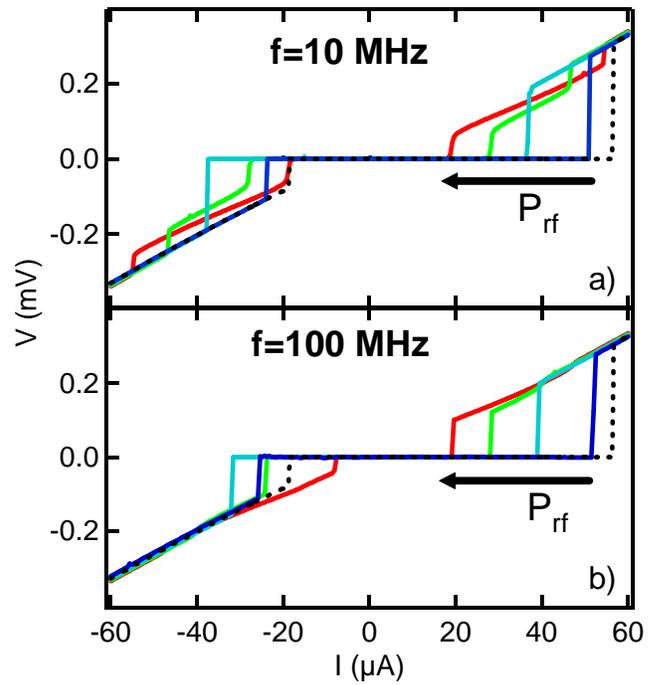}

\caption{(color online) dc voltage vs. dc current measured on sample S with a 10 MHz (a) or 100 MHz (b) rf excitation at several powers : from right to left, P=-66 dBm, -52dBm, -48dBm and -46dBm. The dotted line corresponds to no rf excitation. The temperature is $T=1.5$ K.}

\label{fig_VI}

\end{figure}

\begin{figure}

\includegraphics[width= \columnwidth]{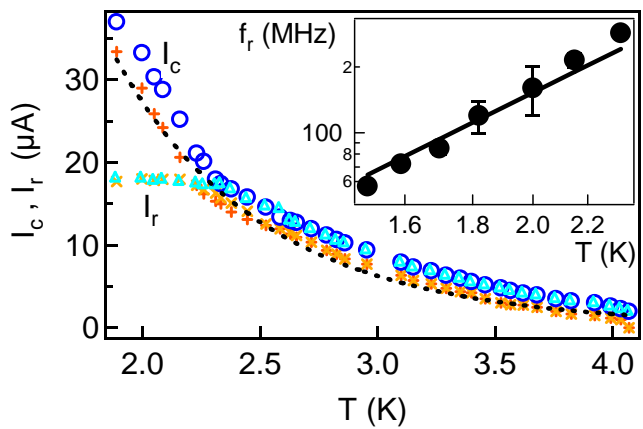}

\caption{(color online) $I_c(T)$ ($+$: rf off, $\bigcirc$: rf on) and $I_r(T)$ ($\times$: rf off, $\bigtriangleup$: rf on) with and without rf excitation (39 GHz, -17.4dBm) on sample S; dotted line: theory for $9.4\times I_c(T)$. Inset: symbols: $f_r$ vs. $T$ for sample S; solid line:  electron-phonon scattering rate for Al vs. $T$.}

\label{fig_IcIrT}

\label{fig_ephT}

\end{figure}

\emph{The low and intermediate frequency regimes --}
A dc $V(I)$ characteristics for sample S at T=1.5K  with no rf excitation is shown on Fig. \ref{fig_VI}, dotted line. One observes a hysteresis (for $T\lesssim2.23$ K, see Fig.\ref{fig_IcIrT}) from which we deduce the critical current $I_c^0$ and the retrapping current $I_r^0$ with no excitation. Let us first consider what happens when a slow ac current of amplitude $I_{ac}$ is added to the dc current. Increasing the dc current $I$ from zero we expect the sample to jump to the N state when $I+I_{ac}=I_c^0$ (for small enough $I_{ac}$). Decreasing $I$ from the N state, the sample stays normal as long as $I-I_{ac}>I_r^0$. Thus for $I_{ac}<I_c^0-I_r^0$, one expects one jump at $I_c=I_c^0-I_{ac}$, i.e. the apparent critical current $I_c$ decreases with increased rf power. When $I_{ac}>I_c^0-I_r^0$, the sample will cycle from S to N and vice-versa, as long as $I_c^0-I_{ac}<I<I_r^0+I_{ac}$. Thus in this regime the $V(I)$ characteristic should have two steps at $I_c^1=I_c^0-I_{ac}$ and $I_c^2=I_r^0+I_{ac}$. The two regimes with one step for small $I_{ac}$ and then a double step for larger $I_{ac}$ is exactly what we observe at 10 MHz, see Fig.\ref{fig_VI}(a), the position of the steps varying as discussed above. Thus the sample follows adiabatically the 10 MHz excitation. When we increase the excitation frequency,  the double steps disappear, see Fig.\ref{fig_VI}(b). More precisely, the measured critical current still verifies $I_c=I_c^0-I_{ac}$ even at high rf power. Therefore when the sample becomes normal, it never jumps back into the superconducting state even though the current $I(t)$ spends some time below the dc retrapping current $I_r^0$. This dynamical effect can be understood if we suppose that the hysteresis in the $V(I)$ curve is at least partially due to heating of the electrons by the dissipative current, as suggested in \cite{Courtois}: the sample can return to the N state only if its temperature is low enough, i.e. when there is not too much Joule power dissipated by the current, which at low frequency occurs at a current $I=I_r^0<I_c^0$. In our case, if the frequency is larger than the inverse thermalization time of the electrons, the instantaneous current may be below $I_r^0$ but electrons are still hot and the sample is in the normal state. The thermalization of the electrons occurs via phonon emission. In the inset of Fig.\ref{fig_ephT} we present the temperature dependence of the frequency $f_r$ corresponding to the transition from two-step to one-step I(V) curves, as well as that of the electron-phonon scattering rate for Al, $\tau_{e-ph}^{-1}=A T^3$, with A being 2.1 times larger than the theoretical value, in good agreement with previous measuremenst \cite{Prober}.  This strongly suggests that the retrapping is related to energy relaxation by electron-phonon interaction.

\begin{figure}

\includegraphics[width=\columnwidth]{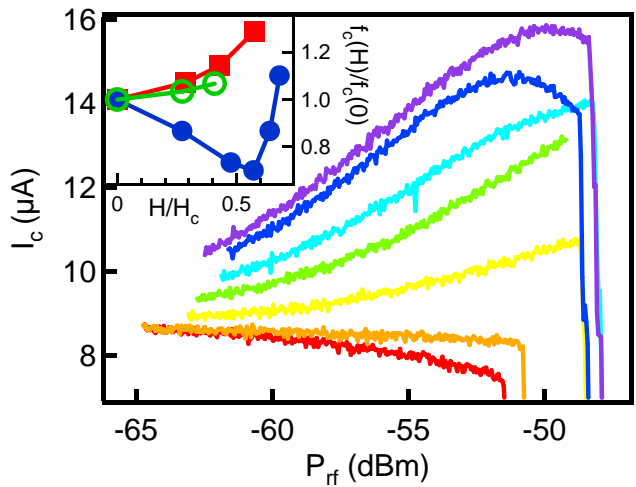}

\caption{(color online) Critical current vs. rf power for different frequencies: from bottom to top f=3,6,12,21,33,36,39 GHz. Inset: symbols: renormalized cross-over frequency $f_c(H)/f_c(0)$ vs. renormalized magnetic field $H/H_0$, where $H_0$ corresponds to one flux quantum in the N part. Full squares: sample L at $T=1.6$ K; full dots: sample S at $T=1.4$ K; circles: sample S at $T=2$ K. }

\label{fig_IcP}

\label{fig_f*H}

\end{figure}

\emph{The high frequency regime --}
We will now consider the high frequency regime. The $V(I)$ characteristics has only single jumps. To measure $I_c$ we current-bias the junction with an 83 Hz linear ramp and measure the time spent before dissipation. In Fig. \ref{fig_IcIrT} we show $I_c(T)$ and $I_r(T)$ for sample S with and without ac excitation (39 GHz, -17.4dBm). We observe that $I_c$ is strongly increased by the rf excitation over the whole temperature range. This is similar to what has been observed in Pb/Cu/Pb junctions \cite{Warlaumont} but in strong contrast with what is observed in superconducting strips, where the enhanced superconductivity is restricted to a very narrow region around $T_c$, of order of a few mK for Al.
The dotted curve of Fig. \ref{fig_IcIrT} corresponds to $\sim9.4$ times the theoretical expectation for $I_c(T)$ in a long SNS junction \cite{Dubos} (there are no adjustable parameters). We think that the factor 9.4 is due to the non zero superconducting coupling constant in the Al.
We will now focus our discussion on the critical current and discuss the retrapping current later. We report in Fig. \ref{fig_IcP} the effect of the rf current on $I_c$ for various frequencies. We observe that below a certain frequency $f_c$ the critical current is decreased by the rf excitation, whereas for $f>f_c$,  $I_c$ is \emph{increased}. We observe an enhancement of $I_c$ by almost a factor two on L samples irradiated at 39 GHz. $I_c$ varies continuously vs. rf power (monotonically up to 33 GHz) up to a threshold at which the sample becomes suddenly normal. This behavior is strikingly different from what has been reported in Sn/Au/Sn junctions and attributed to phase effects, where the increased critical current  smoothly oscillates as a Bessel function of the rf voltage \cite{Mercereau}. Our S and L samples show a similar behavior of $I_c$ vs. rf power, but with a frequency $f_c$ that is larger for short samples: $f_c\sim7$ GHz $\sim0.9\tau_D^{-1}$ for L samples and $f_c\sim17$ GHz $\sim0.4\tau_D^{-1}$ for S samples. In ref. \cite{Warlaumont} is mentioned that the relevant time scale may be the ``effective time-dependent Ginzburg-Landau relaxation time" $\tau^*$ proportional to the diffusion time: $\tau^*=(\pi/2)^2\tau_D$.

In order to elucidate the mechanism that sets the value of $f_c$, we have: i) varied the temperature, and ii) applied a magnetic field $H$ perpendicular to the substrate. We observe: i) no noticeable effect of the temperature on $f_c$, and ii) L and S samples show a non-monotonic behavior of $f_c(H)$, the field dependence being very temperature dependent, see inset of Fig.\ref{fig_f*H}. This suggests a competition between the field dependence of the mini-gap and the magnetic field induced broadening in the quasiparticle density of states.

\emph{The retrapping current --}
We now discuss the effect of the rf excitation on the retrapping current. We show in Fig. \ref{fig_IcIrT} the temperature dependence of $I_r$ with and without rf excitation. One clearly distinguishes two temperature ranges: at high temperature, $T>T^*\sim2.23$ K, there is no hysteresis in the $V(I)$ characteristic, so $I_r=I_c$ is enhanced by the rf excitation. For $T<T^*$, the value of $I_r$ saturates and is barely modified by the rf excitation. Thus, as soon as $I_r$ is a relevant quantity (i.e. when $V(I)$  is hysteretic), its response to the rf excitation is radically different from that of $I_c$. We obtain similar results for samples S with $T^*\sim1.6$ K.

The hysteretic behavior of SIS (I=insulator) Josephson junctions comes from the phase dynamics of the junction that is highly sensitive to the electromagnetic environment. The phase dynamics is usually described by the RSCJ model \cite{Likharev}. Following this model and substituting the characteristic (voltage relaxation) time, $RC$,  by the diffusion time $\tau_D$ \cite{Song,Warlaumont_APL}, one expects $I_r\propto\sqrt{I_c}$, in contradiction with our measurement. Another source of hysteresis in SNS junction is of thermal origin: when in the normal state the sample dissipates Joule power $R_NI^2$. Because of the superconducting contacts, the electrons can relax their energy only by phonon emission whose cooling power is $P_{e-ph}\propto T_e^5-T^5$, with $T_e$ the electron temperature and $T$ the phonon bath temperature. To switch back in the  superconducting state, the current and thus the electron temperature, have to be such that $I<I_c(T_e)$. At high temperature the cooling is very efficient and $I_r\simeq I_c$. At low temperature the cooling is poor, with $T_e$ almost independent of $T$, and $I_r$ saturates to a value significantly lower than $I_c$. This qualitatively reproduces our measurement. However, from our measurement of $I_c(T)$ we find that $I_r$ should saturate at $\sim7\;\mu$A (i.e. $T_e=2.9$ K). This value is 2.5 times smaller than what we measure. More importantly, $I_r$ should be enhanced by the rf excitation as $I_c$ is, and it is not.

\emph{Conclusion --}
SNS systems are intermediate between weakened (constricted) superconductors (Ss'S geometry, s'=weak superconductor) and Josephson junctions with an insulating barrier, and one is tempted to interpret their behavior in terms of these two limits. In constricted superconductors one is dealing with usual bulk superconductivity except that the populations are driven out of equilibrium by the rf pumping. Enhanced superconductivity in Ss'S occurs very close to $T_c$. In contrast, the effect of rf on a Josephson junction arises through the time dependence of the Josephson phase, which leads e.g. to Shapiro steps\cite{noteShapiro}. Our situation is intermediate and a quantitative analysis requires a careful theoretical study of the interplay between out-of-equilibrium superconductivity and phase dynamics in these junctions \cite{Cuevas}. However, using ac irradiation  we performed a photon-assisted like experiment where we have identified two different timescales for the critical and retrapping currents. Specifically, we showed that the critical current and the retrapping current are associated with elastic and inelastic scattering, respectively.

We are grateful to J.-C. Cuevas for showing us some of his results prior to publication.We thank H. Bouchiat, M. Ferrier, J. Gabelli, S. Gu\'eron, I. Petkovic for fruitful discussions and I. Petkovic for technical assistance in the sample preparation. This work was supported by ANR-08-BLAN-0087-01.


\begin{thebibliography}{99}
\bibitem{noteShapiro} We observe Shapiro steps up to 39 GHz in our samples.
\end{thebibliography}
\end{document}